\documentclass{elsart3p}

\journal{Colloids Surfaces B: Biointerfaces}

\usepackage{graphicx}
\usepackage{subfigure}

\begin{document}

\begin{frontmatter}



\title{Electroformation in a flow chamber with solution exchange as a
  means of preparation of flaccid giant vesicles}


\author{Primo\v{z} Peterlin\corauthref{cor}},
\corauth[cor]{Corresponding author. Fax: +386-1-4315127.}
\ead{primoz.peterlin@biofiz.mf.uni-lj.si}
\author{Vesna Arrigler}
\ead{vesna.arrigler@biofiz.mf.uni-lj.si}

\address{University of Ljubljana, Faculty of Medicine, Institute of
Biophysics, Lipi\v{c}eva 2, SI-1000 Ljubljana}

\begin{abstract}
  A recently described technique (Estes and Mayer,
  Biochim. Biophys. Acta 1712 (2005) 152--160) for the preparation of
  giant unilamellar vesicles (GUVs) in solutions with high ionic
  strength is examined.  By observing a series of osmotic swellings
  followed by vesicle bursts upon a micropipette transfer of a single
  POPC GUV from a sucrose solution into an iso-osmolar glycerol
  solution, a value for the permeability of POPC membrane for
  glycerol, $P = (2.09\pm0.82)\cdot 10^{-8}\,\textrm{m/s}$, has been
  obtained.  Based on this result, an alternative mechanism is
  proposed for the observed exchange of vesicle interior.  With
  modifications, the method of Estes and Mayer is then applied to
  preparation of flaccid GUVs.
\end{abstract}

\begin{keyword}
  giant vesicle \sep electroformation \sep conductive solution \sep
  membrane permeability \sep glycerol

\end{keyword}
\end{frontmatter}

\section{Introduction}
\label{sec:intro}

Since being introduced over 20 years ago by Angelova and Dimitrov
\cite{Angelova:1986}, the electroformation method has became an
indispensible tool for preparing giant unilamellar vesicles (GUVs)
with a diameter exceeding 10~$\mu$m, which constitute useful model
systems for biological cells \cite{luisi:GiantVesicles:2000}.  Being
first developed with a DC electric field \cite{Angelova:1986}, the
method was subsequently improved by replacing the DC electric field
with a very low-frequency (1--10~Hz) AC electric field
\cite{Angelova:1987}.  Two experimental setups were established: in
the first one a pair of parallel platinum wires constituted the
electrodes \cite{Angelova:1988}, while the other used plan-parallel
electrodes instead: microscope slides, coated with an optically
transparent layer of indium tin oxide (ITO) \cite{Angelova:1992}.  An
overview of the development is available in \cite{Angelova:2000}.
Recently, a number of techniques have been developed which combine
electroformation with other techniques, including micropatterning of
ITO glass \cite{Taylor:2003b}, microfluidic channels
\cite{Estes:2006,Kuribayashi:2006}, and electroformation on
non-conductive substrates \cite{Okumura:2007}.

While the electroformation method is appreciated for its superior
homogeneity of vesicle sizes \cite{Bagatolli:2000} and is widely
recognized as a simple and reproducible technique which appears to be
efficient for many types of different lipid mixtures
\cite{Fischer:2000}, its key disadvantage is that it cannot be applied
if the salt concentration exceeds 10~mM \cite{Mathivet:1996}, or when
the membrane is composed of charged amphiphiles \cite{Rodriguez:2005}.
Methods based on spontaneous swelling \cite{Reeves:1969,Needham:1988}
were used for the preparation of GUVs in physiological conditions
\cite{Akashi:1996,Moscho:1996,Yamashita:2002}.  Recently, however,
Estes and Mayer proposed a technique where electroformation is
conducted in a flow chamber \cite{Estes:2005b}: in the first step,
electroformation is performed as usual, and in the second step the
solution in the electroformation chamber is exchanged while the
vesicles are still attached to the electrodes.  In this way, the
authors report to be able to produce GUVs (10--100~$\mu$m in diameter)
in solutions with high ionic strength (up to 2~mol/L KCl).  Employing
phase contrast and fluorescence techniques, they report that within 30
minutes, the solution \emph{inside} the vesicles has also been
replaced, and propose a diffusion mechanism through lipid tubules to
explain this phenomenon.

In this paper, we examine the technique of Estes and Mayer with
different pairs of internal/external solutions.  We estimate the
permeability of POPC membrane for glycerol with an experiment, where a
single POPC GUV is transferred with a micropipette from a sucrose
solution into an iso-osmolar glycerol solution.  Based on this
relatively high permeability, we provide an alternative explanation
for the solution exchange inside the vesicles.  We also extend the
work of Estes and Mayer and apply their method to demonstrate the
possibility of its use for the preparation of flaccid GUVs.

\section{Materials and methods}
\label{sec:materials-methods}

\subsection{Materials}
Trizma base, Trizma HCl, and L-$\alpha$-phosphatidyl\-choline from egg
yolk (eggPC) were purchased from Sigma-Aldrich (St. Louis, USA);
D-(+)-glucose, D-(+)-sucrose and glycerol were from Fluka (Buchs,
Switzerland).  Methanol and chloroform were purchased from Kemika
(Zagreb, Croatia).
1-palmitoyl-2-oleoyl-\textit{sn}-glycero-3-phosphocholine (POPC) was
purchased from Avanti Polar Lipids (Alabaster, USA).
2-(12-(7-nitrobenz-2-oxa-1,3-diazol-4-yl)amino)do\-deca\-noyl-1-hexadecanoyl-sn-glycero-3-phosphocholine
(NBD C${}_{12}$-HPC) was purchased from Invitrogen (Eugene, USA).  All
the solutions were prepared in double-distilled and sterile water.

\subsection{Electroformation chamber and procedure}
\label{sec:procedure}

The electroformation chamber was machined from two plates of 2.5~mm
acrylic glass (polymethyl methacrylate, PMMA), mounted together with
four M3 screws.  A $20\times 12$~mm opening was cut through both
plates, yielding an observation chamber with a volume of
1.2~$\textrm{cm}^3$.  A channel 1.5~mm deep and 2.5~mm wide was milled
into the inner face of both plates from two diagonally opposite
corners of the observation chamber to the edge, allowing for mounting
the inlet and the outlet tube (PVC perfusor tube, outside diameter
2~mm; Tik, Kobarid, Slovenia).  Another channel 1~mm deep and 1.5~mm
wide was milled into the inner face of the smaller plate to allow for
the access of electrodes made of 1~mm Pt wire, which were also mounted
with M3 screws.  The distance between the electrodes is 5~mm. The
observation chamber is covered with a $32 \times 24$~mm glass cover
slip on the upper and the lower side, mounted with a vacuum grease
(Baysilone; Bayer, Leverkusen, Germany); vacuum grease is also used to
seal the contact surface between the two plates and the openings for
tubing and the electrodes.

\begin{figure}
  \centering\includegraphics[scale=0.8]{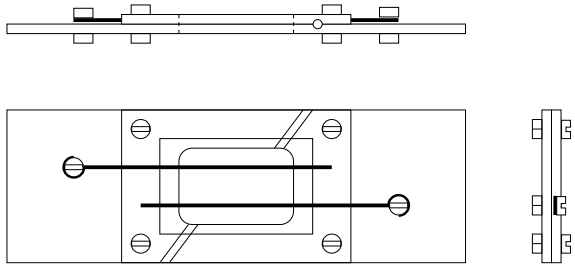}
  \caption{The electroformation chamber is machined from two pieces
  of 2.5~mm acrylic glass, mounted together with four M3 screws. The
  drawing is not to scale.}
  \label{fig:electroform-chamber}
\end{figure}

A suspension of POPC GUVs (\emph{e.g}, in 0.2~mol/L sucrose/glucose
solution, 1:1, or in mixtures of other iso-osmolar solutions) was
prepared using an electroformation method described by Angelova
\cite{Angelova:1988} with modifications
\cite{Heinrich:1996,Mally:2002}.  The lipids were dissolved in a
mixture of chloroform/methanol (2:1, v/v) to a concentration
1~mg/mL. 25~$\mu$L of the lipid solution was spread onto two Pt
electrodes and dried under the reduced pressure (water aspirator;
$\approx 60$~mmHg) for 2~hours.  The electrodes were then placed into
an electroformation chamber, which was filled with 0.2~mol/L sucrose.
AC current (8~V, 10~Hz) was then applied. After 2~hours the voltage
and the frequency were reduced in steps, first to 4~V/5~Hz, after
15~minutes to 2.5~V/2.5~Hz, and after additional 15~minutes to the
final values of 1~V and 1~Hz, which were held for 30~minutes.  From
this point, either of the two procedures were employed: (a) In a
flow-exchange experiment, the syringe pump was switched on at this
stage; (b) If a suspension of vesicles in 0.2~mol/L sucrose/glucose
solution was prepared for the micropipette experiment, the chamber was
drained into a beaker at this stage and flushed with buffered
0.2~mol/L glucose solution, thus resulting in a suspension of GUVs in
a 1:1 sucrose/glucose solution.  This procedure yields preferentially
spherical unilamellar vesicles with diameters up to 100~$\mu$m
containing entrapped sucrose.

\subsection{Flow exchange}

A syringe pump (Genie; Kent Scientific Corp., Torrington, USA) was
used to exchange the solution in the flow chamber at a constant flow
rate.  A volumetric flow rate set to $\Phi_V = 5$~mL/h resulted in a
movement of fluid past the electrodes, its speed estimated at $v\sim
30$~$\mu$m/s.  An estimate for the Reynolds number of the solution
flow past the electrodes yields $\textrm{Re} = l \rho v/\eta \sim
0.03$, with $l \sim 1$~mm being the characteristic length (electrode
diameter), $\rho \approx 1$~kg/L the density of the solution, and
$\eta \approx 8.9\cdot 10^{-4}$~$\textrm{Pa\,s}$ the dynamic viscosity
of the solution, which we approximated with the value for water at
25${}^\circ$C.  This means the flow in the chamber can be adequately
approximated with a laminar flow.

\subsection{Microscopy and micromanipulation}

Both phase contrast and fluorescent micrographs were obtained with an
inverted optical microscope (Nikon Diaphot 200, objective 20/0.40 Ph2
DL or Fluor 60/0.70 Ph3DM) with the epi-fluorescence attachment,
micromanipulating equipment (Narishige MMN-1/MMO-202) and a cooled CCD
camera (Hamamatsu ORCA-ER; C4742-95-12ERG), connected to a PC running
Hamamatsu Wasabi software.  The software also controlled a Uniblitz
shutter with VMM-D1 controller (Vincent Associates, Rochester NY, USA)
in the light path of the Hg-arc light source.

\begin{figure}
  \centering\includegraphics[scale=0.35]{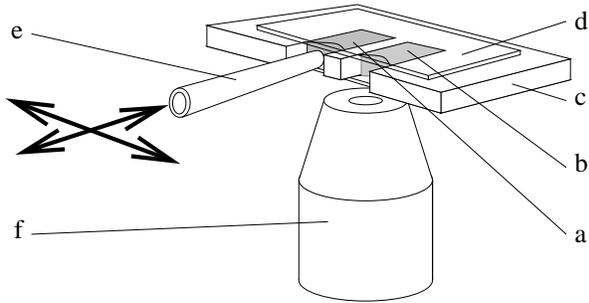}
  \caption{Micromanipulation setup for transferring GUVs into a
    chamber with glycerol solution. a---chamber with the suspension of
    POPC GUVs in a sucrose solution; b---chamber with the glycerol
    solution; c---frame machinned from 2~mm acrylic glass, mounted to
    the microscope stage; d---top $32 \times 24$ mm glass cover slip
    (an identical cover slip is on the bottom side), mounted with a
    vacuum grease; e---micropipette mounted to the micromanipulator;
    f---microscope objective. The scheme is not drawn to scale.}
  \label{fig:micromanipulation}
\end{figure}

In a micromanipulation experiment, a spherical POPC GUV was selected,
fully aspirated into a glass micropipette with a diameter exceeding
the diameter of the vesicle, and transferred into the target solution
of glycerol, where the content of the micropipette was released
(Fig. \ref{fig:micromanipulation}).  After removing the micropipette,
the vesicle was monitored with a video camera. To allow for access
with a micropipette, both compartments in a micromanipulation chamber
are relatively thick (approx. 2.5--3~mm) and open on one side.

\subsection{Refractometry}

Refractive indices of glycerol, NaCl, glucose, sucrose, and buffer
solutions were measured using an Abbe refractometer (Xintian WY1A,
Guiyang, China).

\section{Results}
\label{sec:results}

\subsection{Control experiments}
\label{sec:control}

As a control, the experiment of Estes and Mayer \cite{Estes:2005b} was
repeated in several variations, both with ionic (NaCl, Trizma buffer)
and non-ionic (glycerol, sucrose, glucose) solutions. 

\subsubsection{Glycerol exchange} 
\begin{figure*}
  \begin{center}
    \subfigure[]{\includegraphics[width=0.3\linewidth]{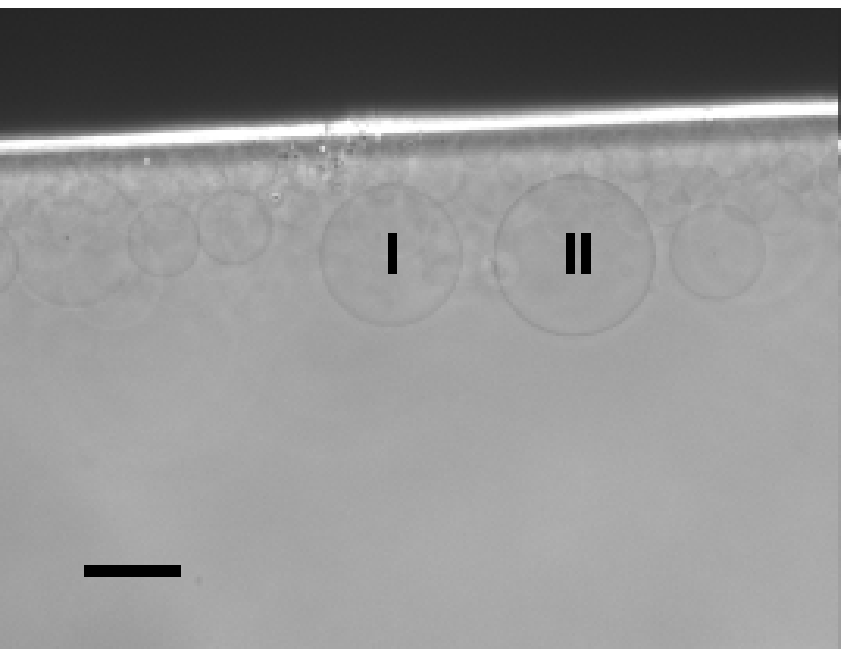}}
    \subfigure[]{\includegraphics[width=0.3\linewidth]{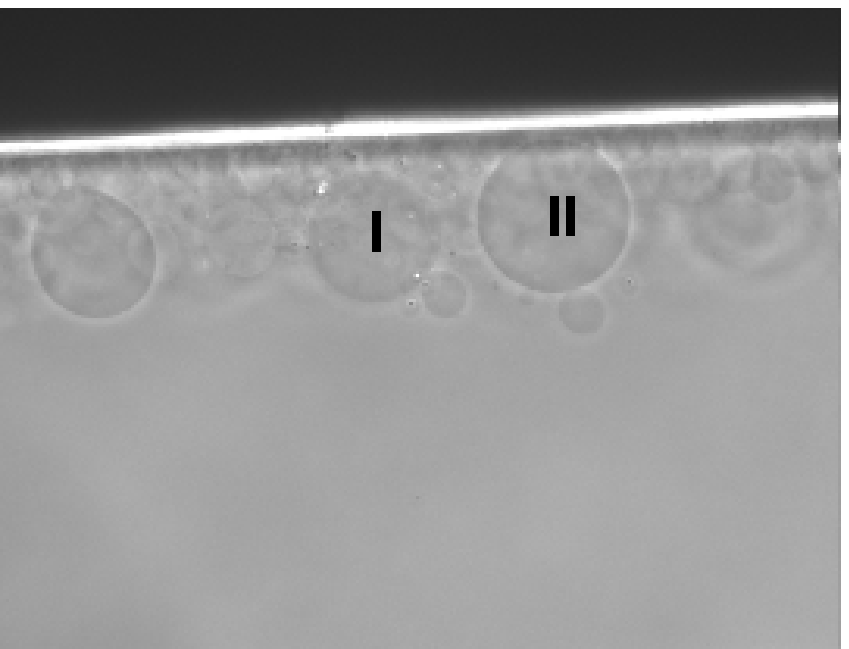}}
    \subfigure[]{\includegraphics[width=0.3\linewidth]{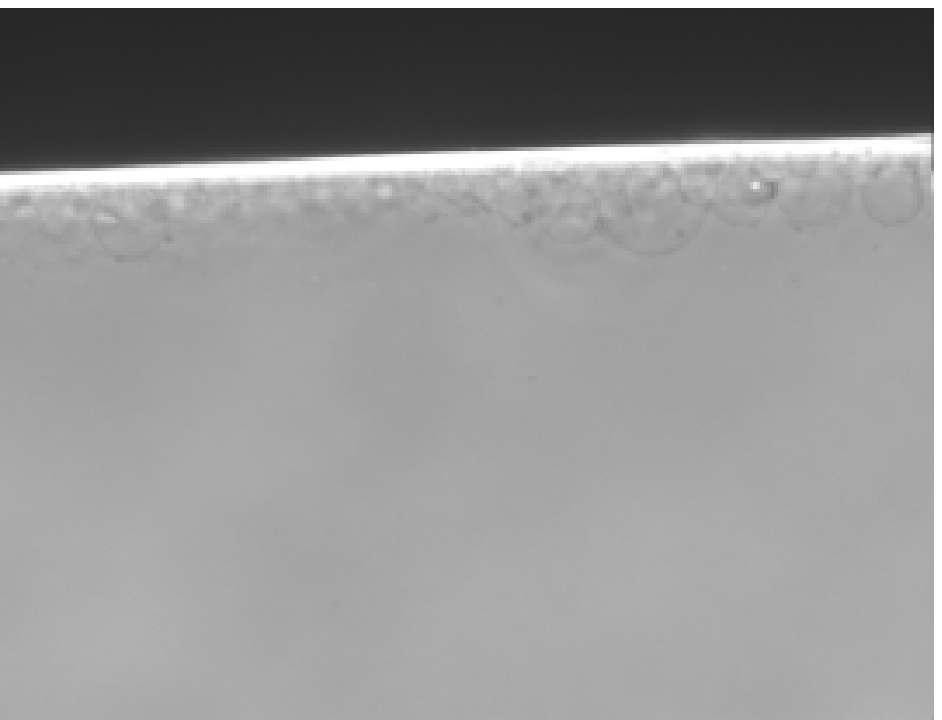}}\\
    \subfigure[]{\includegraphics[width=0.3\linewidth]{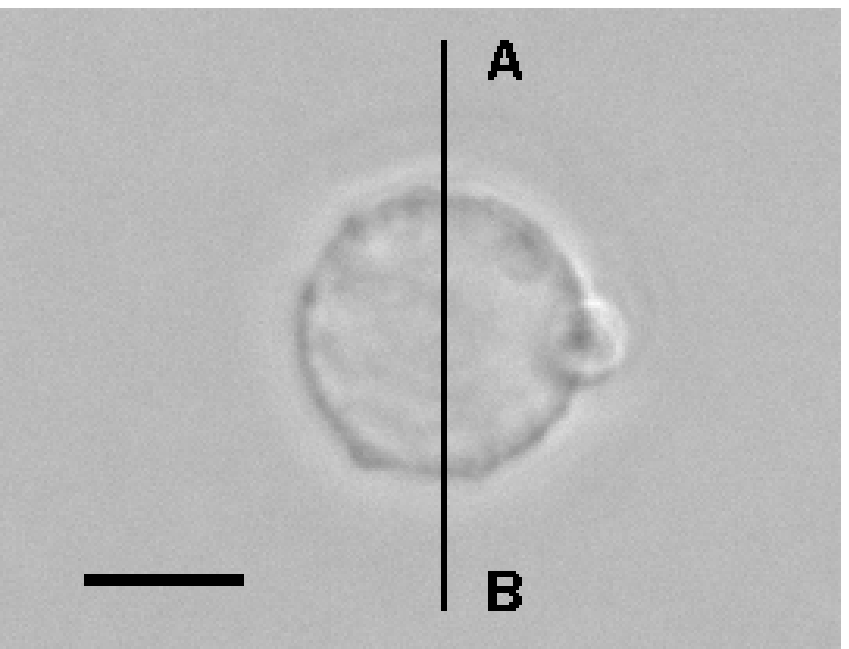}}
    \subfigure[]{\includegraphics[width=0.372\linewidth]{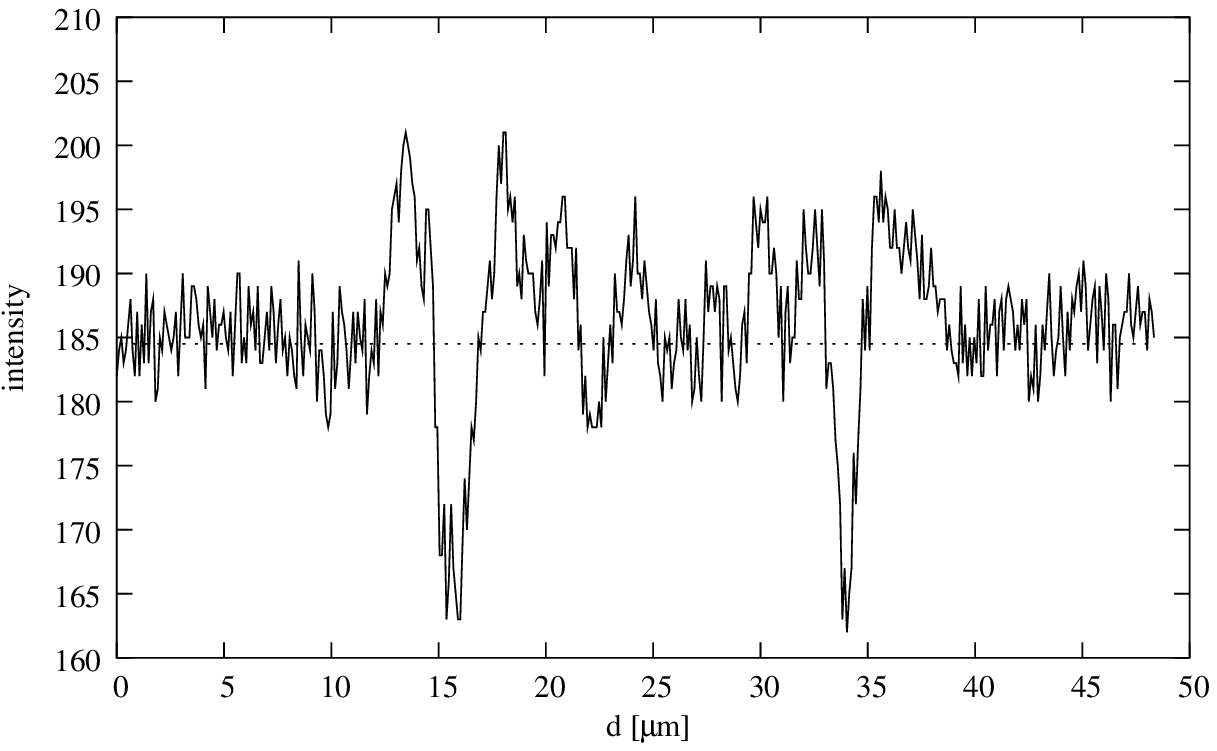}}
  \end{center}
  \caption{Solution exchange in a flow chamber: POPC in 0.2 mol/L
    glycerol substituted with an iso-osmolar 0.1 mol/L NaCl solution.
    (a) before the exchange, (b) 6 minutes, and (c) 48 minutes after
    the NaCl solution reached the flow chamber.  Two vesicles (marked
    I and II) were traced from (a) to (b), while (c) shows another
    section of the electrode.  (d) Vesicles after having been drained
    from the flow chamber.  (e) Intensity profile of the cross-section
    A--B indicated in (d).  The bar in the frame (a) represents 50
    $\mu$m; the photomicrographs in frames (a)--(c) are in the same
    scale.  The bar in the frame (d) represents 10 $\mu$m.}
  \label{fig:exchange-glycerol}
\end{figure*}

In one experiment, we performed electroformation as described above
with either EggPC or POPC in 0.2~mol/L glycerol solution ($n=1.3395$,
26${}^\circ$C), and replaced the solution with an iso-osmolar
0.1~mol/L NaCl solution ($n=1.3378$, 26${}^\circ$C) after the vesicles
were formed.  As reported by Estes and Mayer \cite{Estes:2005b}, three
stages can be distinguished.  Initially, there was the same solution
both inside and outside the vesicles, and vesicle interiors appeared
equally dark as the background (Fig.~\ref{fig:exchange-glycerol}a).
In an intermediate phase, the vesicle interior appears darker in a
phase contract microscopy, and also exhibits a characteristic white
halo around the vesicle (Fig.~\ref{fig:exchange-glycerol}b).  This is
characteristic of a mismatch in the refractive indices of the internal
solution (0.2~mol/L glycerol) and the external solution (0.1~mol/L
NaCl).  The intensity and the width of the halo are proportional to
the difference of refractive indices, and, when properly calibrated,
can be used for a quantitative determination of the concentration in a
vesicle interior \cite{Mally:2002,Mally:2007}.  After some time,
however, the intensity and the width of the halo and the difference in
darkness decrease (Fig.~\ref{fig:exchange-glycerol}c), indicating that
substituting the glycerol solution in the flow chamber with the NaCl
solution resulted in an exchange of the vesicle interior with the
exterior.  Both EggPC and POPC exhibited the same behaviour. As
evident from Figs.~\ref{fig:exchange-glycerol}a--c, increasing ionic
strength causes the aggregation of vesicles and subsequently their
aggregation on the electrodes.  It also shows that the apparent radius
of vesicles decreases with an increasing ionic strength, while
simultaneously the vesicles grow internal spherical invaginations
(visible on Fig.~\ref{fig:exchange-glycerol}d).  We attribute a
relatively low yield of this method---considerably fewer vesicles were
found in the drained suspension than in a conventional
electroformation experiment---to this observed vesicle aggregation on
the electrodes.

Similar three-phase behaviour, indicating that the solution in the
vesicle interior has been at least partly exchanged with the one in
the flow chamber, has been observed with exchanging glycerol with
non-ionic solutions, \emph{e.g.}, when 0.2~mol/L glycerol solution was
substituted with 0.2~mol/L glucose or vice versa (not shown).  The
difference between this experiment and the experiment described above
is that no vesicle aggregation was observed in this case.

\subsubsection{Sugar exchange}
In this set of experiments, glycerol was not used. Instead, we
replaced a non-ionic solution (0.2~mol/L sucrose solution, $n=1.3464$,
26${}^\circ$C or 0.2~mol/L glucose solution, $n=1.3417$, 26${}^\circ$C)
with another non-ionic solution, or with an ionic solution (0.1~mol/L
NaCl, or Trizma buffer). Experiments were conducted with both POPC and
EggPC.

\begin{figure*}
  \begin{center}
    \subfigure[]{\includegraphics[width=0.3\linewidth]{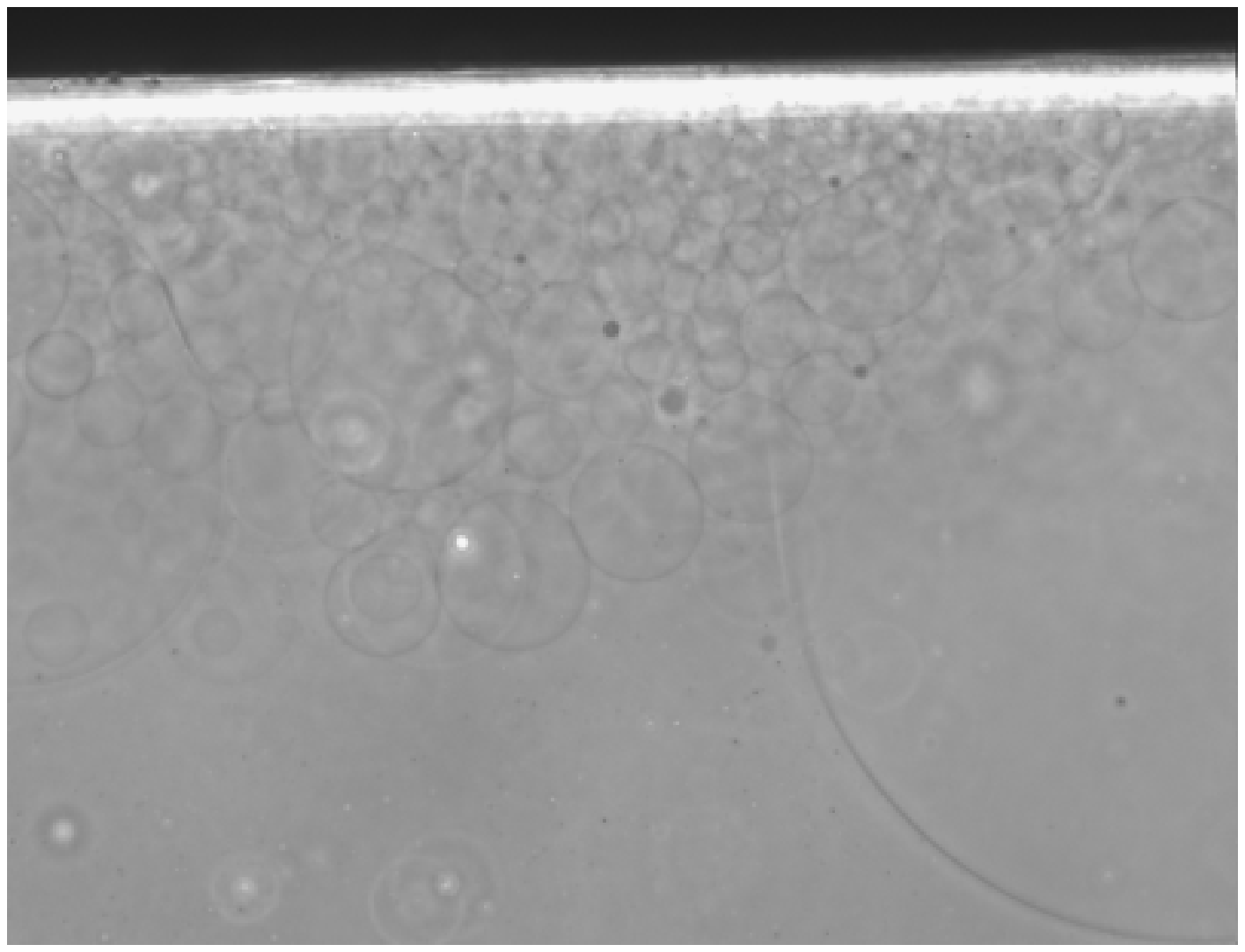}}
    \subfigure[]{\includegraphics[width=0.3\linewidth]{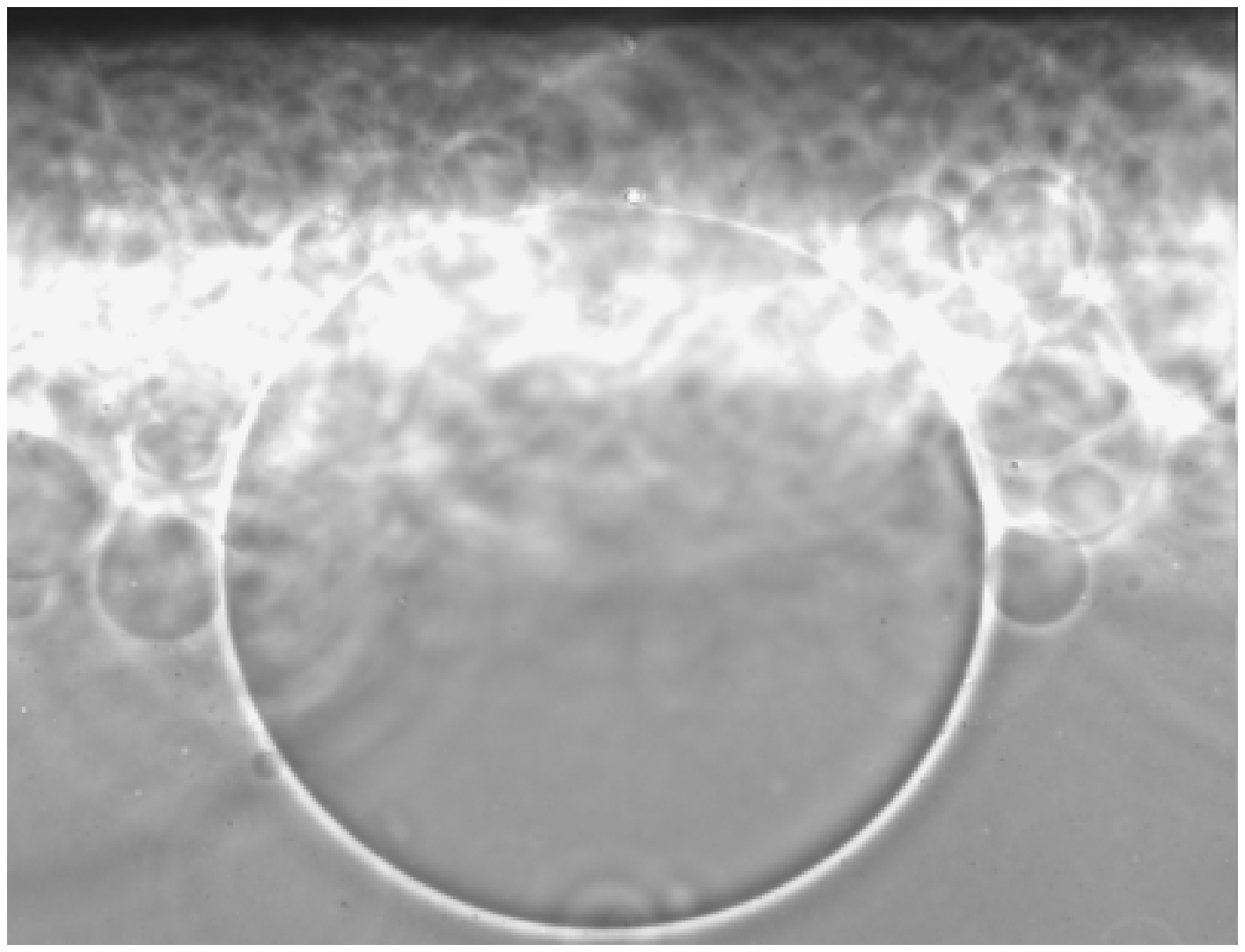}}
    \subfigure[]{\includegraphics[width=0.3\linewidth]{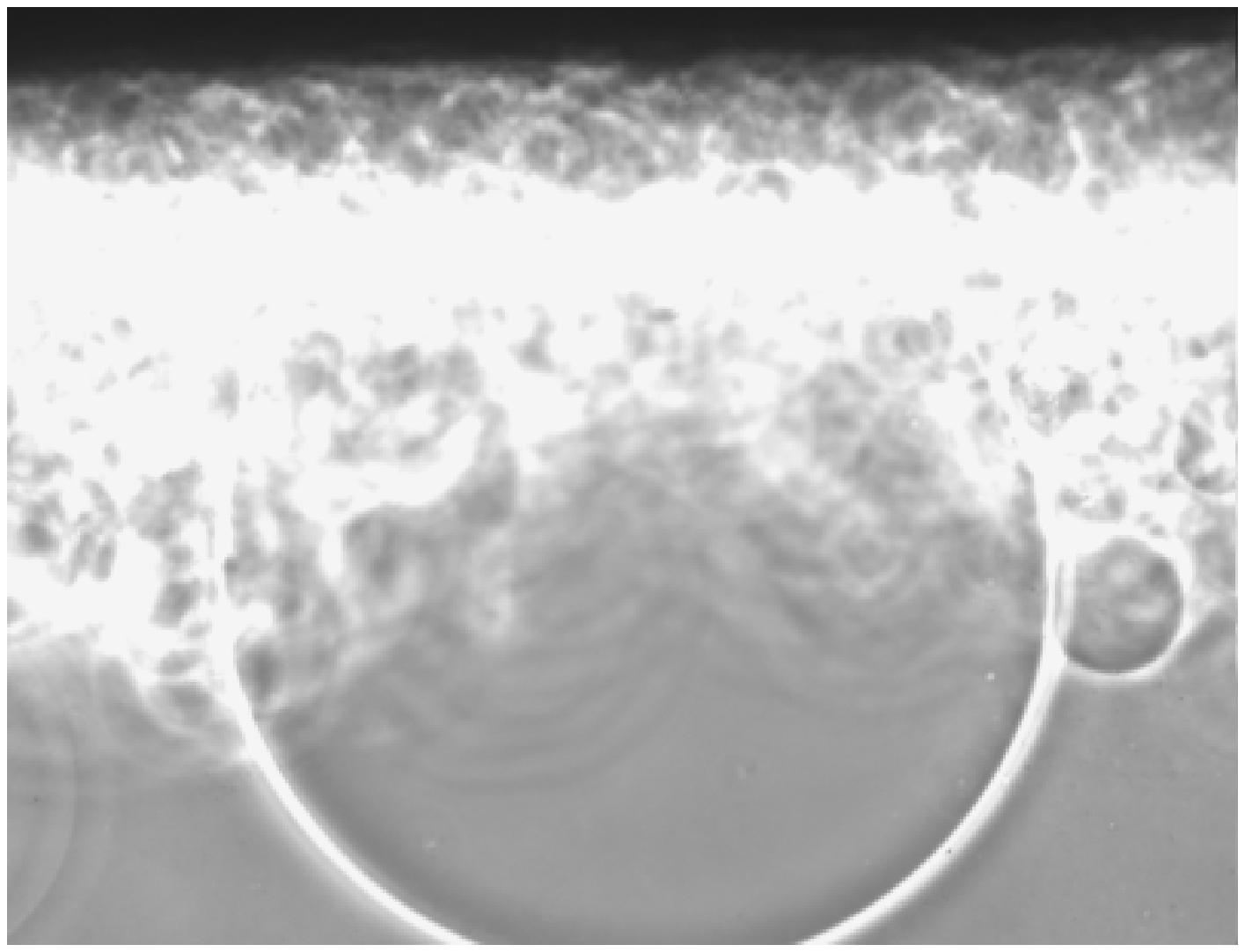}}\\
    \subfigure[]{\includegraphics[width=0.3\linewidth]{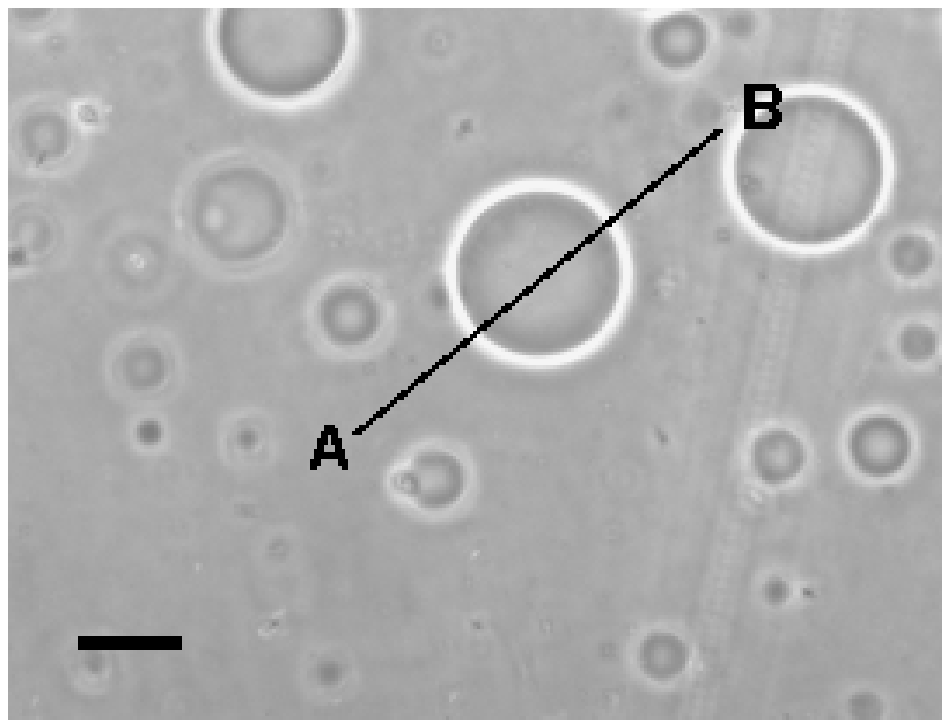}}
    \subfigure[]{\includegraphics[width=0.38\linewidth]{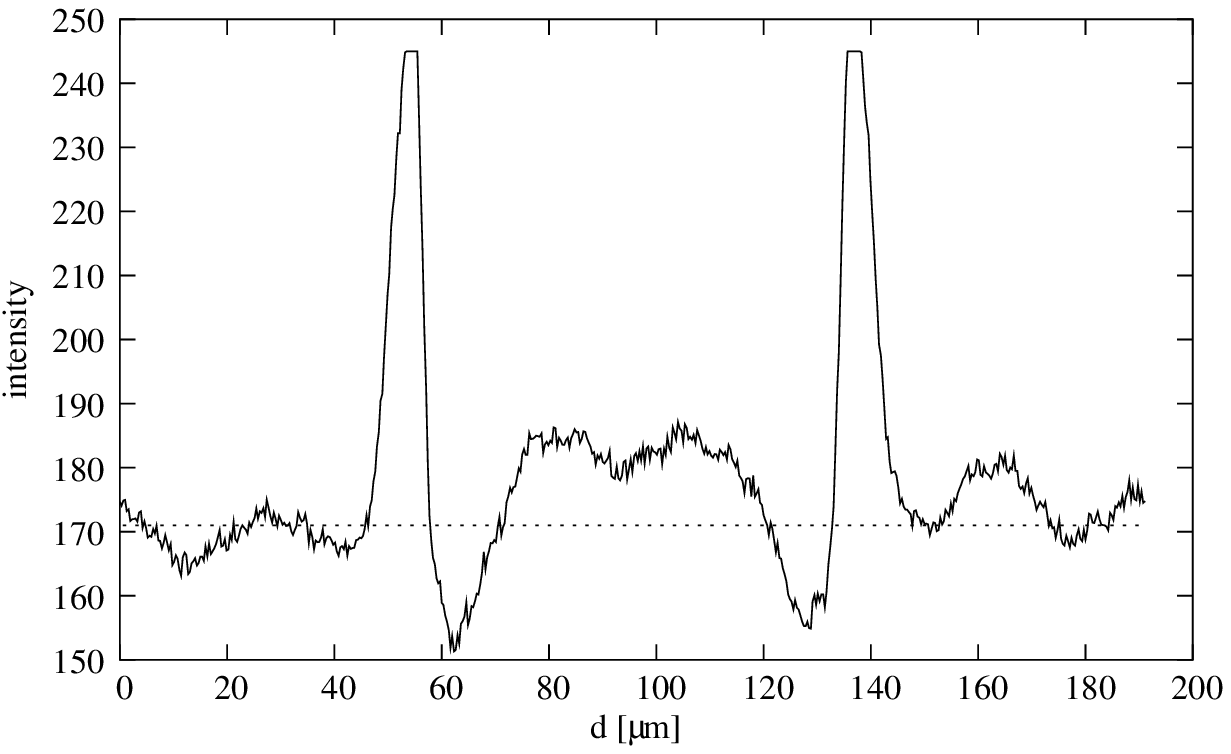}}
  \end{center}
  \caption{Solution exchange in a flow chamber: POPC in 0.2 mol/L
    sucrose substituted with an iso-osmolar 0.1 mol/L Trizma buffer (pH
    9.0). (a) before the exchange; (b) 3 minutes, and (c) 28 minutes
    after the buffer solution reached the flow chamber. (d) Vesicles
    after having been drained from the flow chamber. (e) Intensity profile
    of the cross-section A--B indicated in (d). The bar in the frame
    (d) represents 50 $\mu$m; the photomicrographs in the frames
    (a)--(d) are in the same scale.}
  \label{fig:exchange-sucrose}
\end{figure*}

In those experiments which did not involve glycerol solution
(Fig.~\ref{fig:exchange-sucrose}), we only observed two phases: we
started with the same solution inside and outside the vesicle
(Fig.~\ref{fig:exchange-sucrose}a), then, as the external solution was
exchanged, we noticed a change in darkness of the vesicle interior and
exterior, and the appearance of the characteristic halo around a
vesicle (Fig.~\ref{fig:exchange-sucrose}b,c). The halo did not
significantly diminish with time.

The same two-phase behaviour without an exchange of vesicle interior
was noticed with EggPC and POPC vesicles when 0.2~mol/L sucrose
solution was substituted with 0.1~mol/L NaCl, and with POPC vesicles
when 0.2~mol/L sucrose solution was exchanged with an iso-osmolar
solution, either non-ionic (0.2~mol/L glucose), or ionic (0.15~mol/L
Trizma buffer, pH~7.4, $n=1.3397$, or 0.1~mol/L Trizma buffer, pH~9,
$n=1.3397$). The ionic and non-ionic solutions differed in the fact
that vesicle aggregation was only noticed in ionic solutions.

\subsubsection{Solution exchange during electroformation}

\begin{figure*}
  \begin{center}
    \subfigure[]{\includegraphics[width=0.3\linewidth]{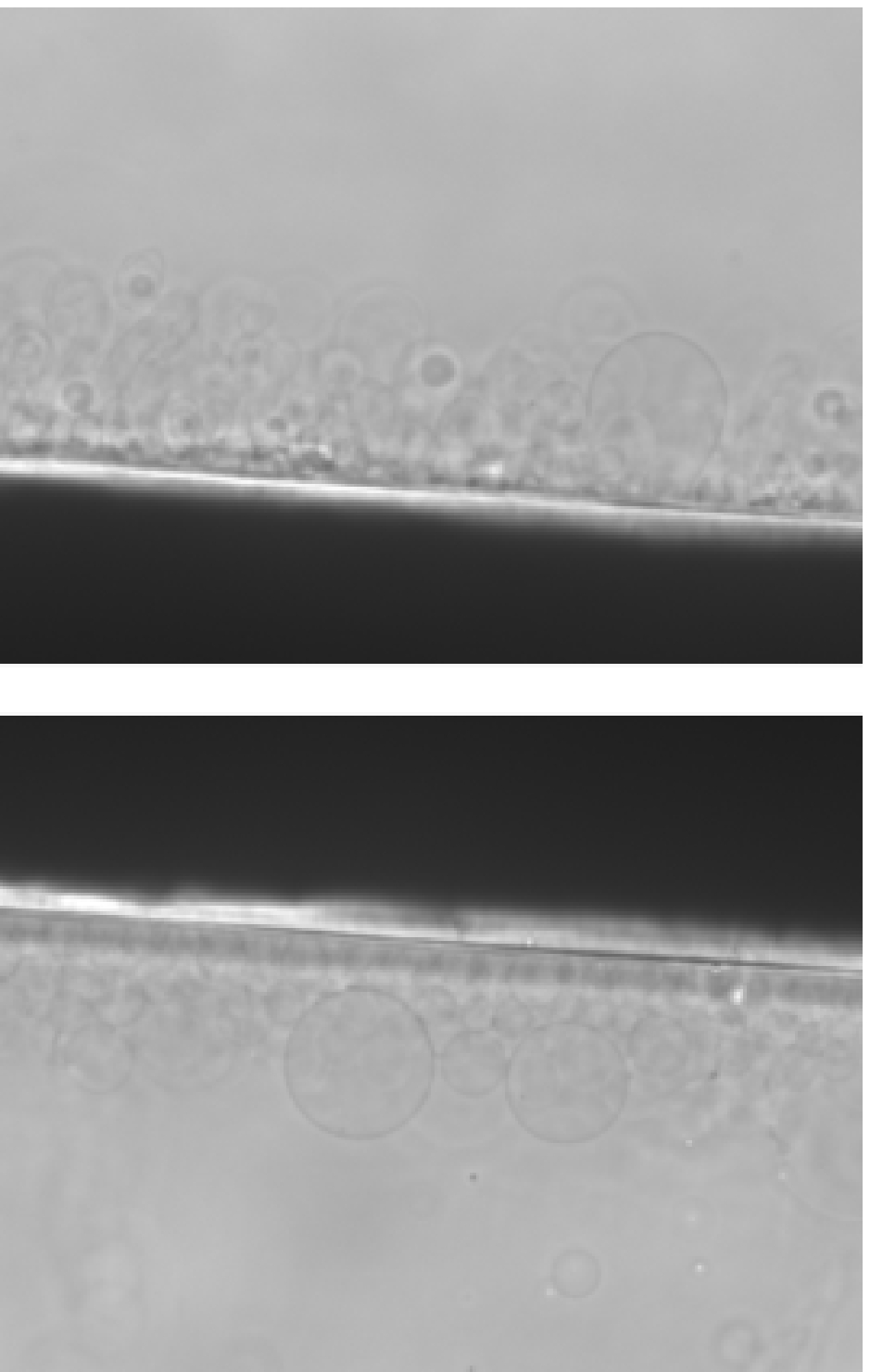}}
    \subfigure[]{\includegraphics[width=0.3\linewidth]{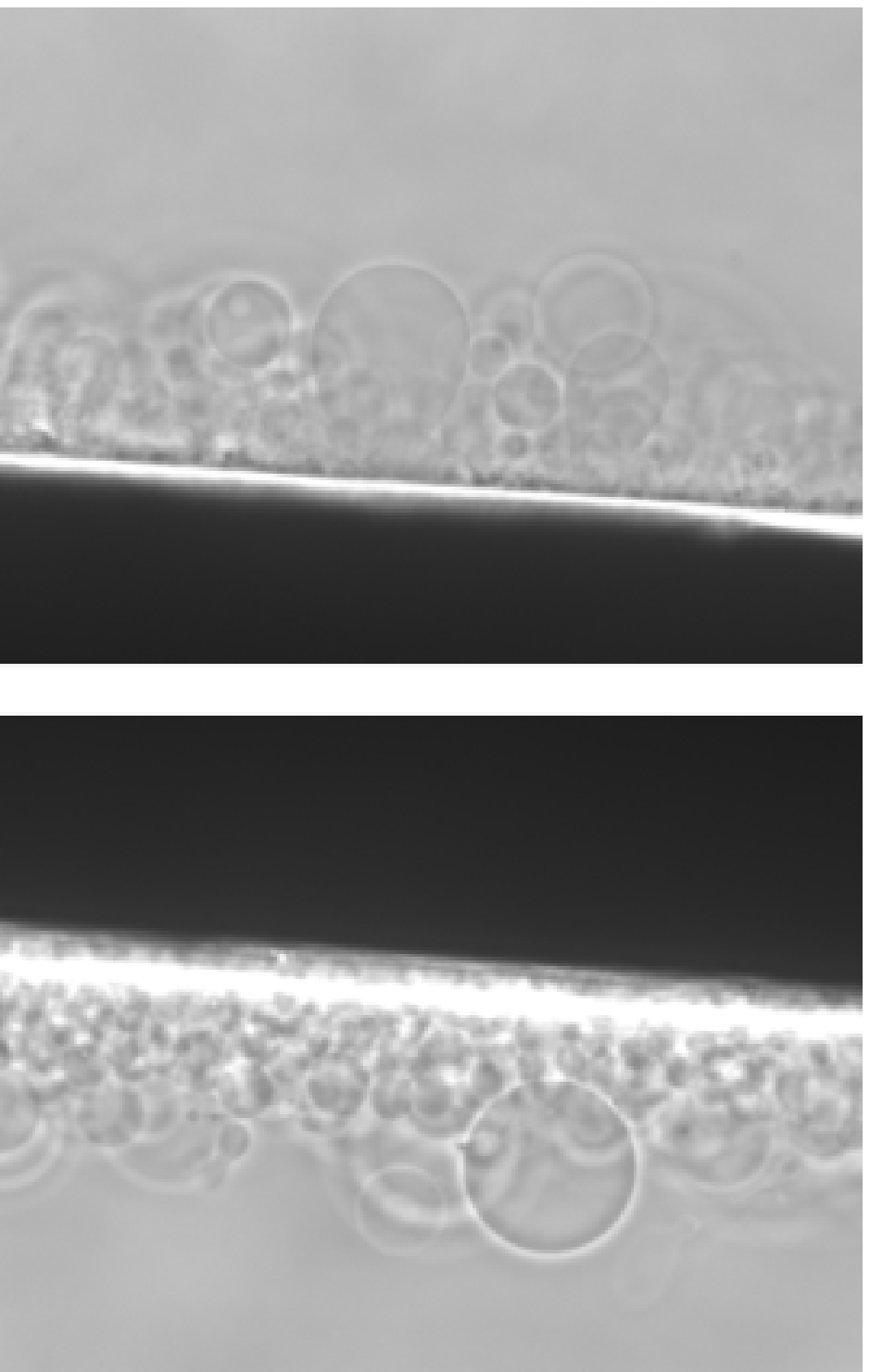}}
    \subfigure[]{\includegraphics[width=0.3\linewidth]{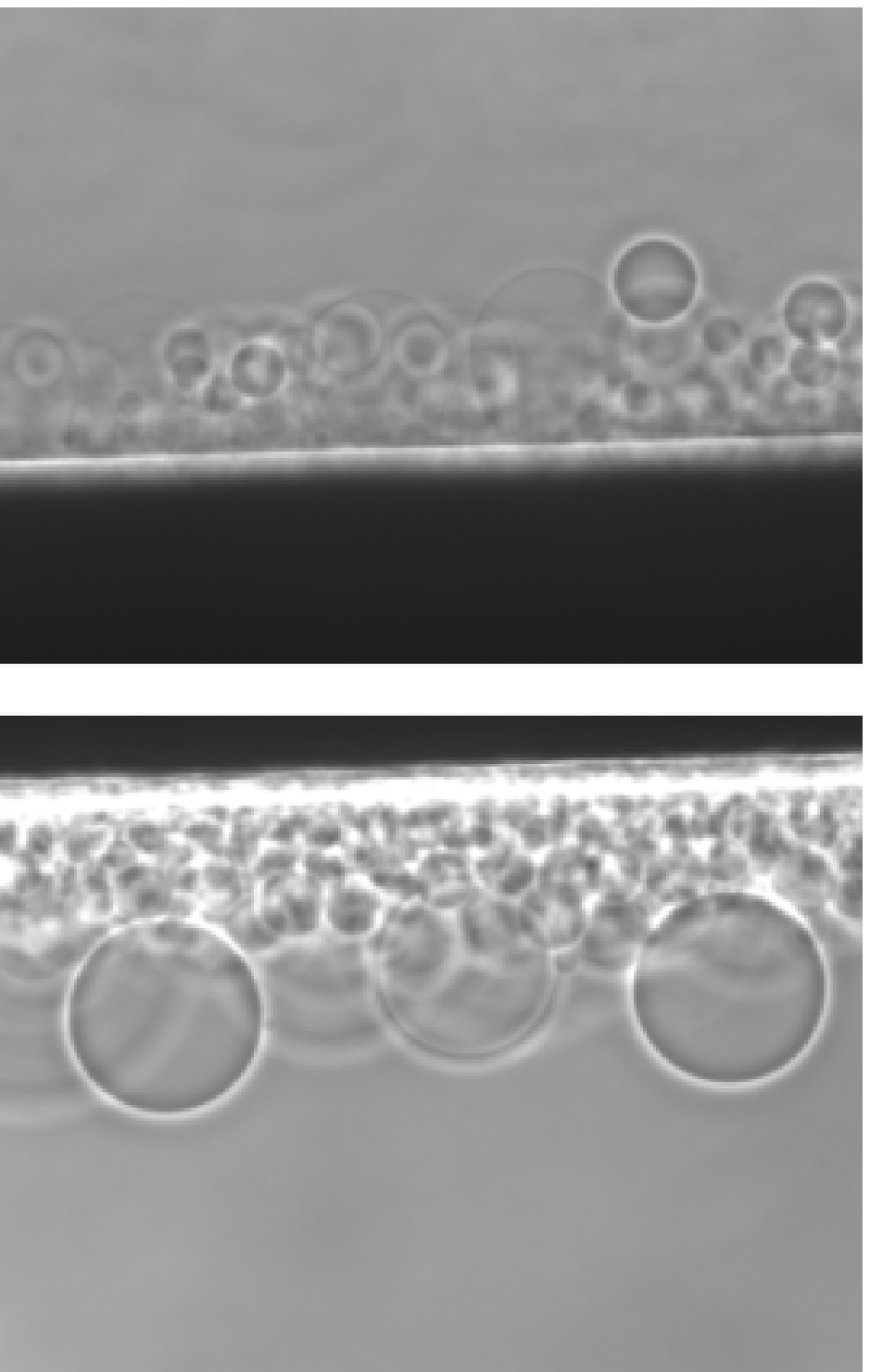}}
  \end{center}
  \caption{Vesicles on the electrodes when the solution was exchanged
    during electroformation. POPC in 0.2~mol/L sucrose is substituted
    by 0.2~mol/L glucose.  (a) Vesicles on the outer (top) and the
    inner (bottom) side of the electrode 11 minutes (bottom)/11 minuts
    40~s (top); (b) 22~min 30~s (top)/23~min 10~s (bottom), and (c)
    37~min 50~s (top)/39~min 10~s after the simultaneous application
    of the electric field and the start of the flow.  The bar in the
    frame (a) represents 50 $\mu$m, all photomicrographs are in the
    same scale.}
  \label{fig:exchange-during}
\end{figure*}

In this experiment the solution exchange was initiated shortly---at
the given length of the inlet tube and the flow rate, it took $23\pm
1$ minutes for the replacement solution to reach the electroformation
chamber in our setup, as verified independently---after the
electroformation was started.  In the experiment where 0.2~mol/L
sucrose solution was substituted with 0.2~mol/L glucose solution, we
found that this time was sufficient for the entrapment of the vesicle
interior.  As visible in Fig.~\ref{fig:exchange-during}a, 11~minutes
from the onset, spherical vesicles are already formed on the inner
side of the electrode surface (bottom), while the outer side (top) is
predominantly populated by tubular structures with wider ``necks''.
23~minutes from the onset, glucose has reached the electroformation
chamber, as verified in Fig.~\ref{fig:exchange-during}b, where a weak
halo can be seen around the vesicles both on the inner and on the
outer side of the electrode.  In particular, the halo is also visible
around the tubular membrane structures on the outer side of the
electrode (Fig.~\ref{fig:exchange-during}b, top).  38~minutes from the
onset, the solution in the vesicle exterior has been completely
exchanged, as verified by the halo around the vesicles; the intensity
of the halo has reached a plateau (Fig.~\ref{fig:exchange-during}c).
The inner side of the electrode is populated by spherical vesicles
surrounded by a strong halo (bottom), while on the outer side,
spherical vesicles are smaller in size. There is a noticeable halo
around spherical vesicles, while there is no halo around the tubular
membrane structures with wider ``necks'', meaning that their interior
has the same composition as the exterior.  In these cases, the
exchange has likely taken place via the mechanism proposed by Estes and
Mayer \cite{Estes:2005b}.

The difference between the vesicle growth rates on the inner and on
the outer electrode surface indicates a difference in the electric
field strength affecting the adsorbed lipid layer.  An exact estimate
of the field, however, is not trivial to obtain (see
appendix~\ref{sec:bipolar} for an estimate and a discussion).

\subsection{Micropipette transfer into glycerol}

\begin{figure}
  \centering\includegraphics[scale=0.6]{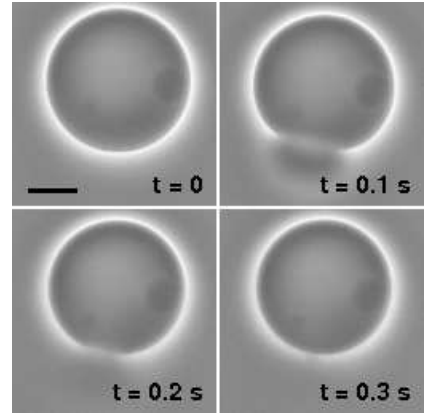}
  \caption{Burst after a micropipette transfer of a POPC GUV from
    0.2~mol/L sucrose/glucose solution into 0.2~mol/L glycerol
    solution. The bar in the top left frame represents 20~$\mu$m.}
  \label{fig:burst}
\end{figure}

In this experiment, fully formed POPC GUVs were transferred using a
micropipette into an iso-osmolar glycerol solution.  POPC GUVs were
prepared in 0.2~mol/L sucrose/glucose solution as described above and
left overnight. Within 2 minutes after the transfer into 0.2~mol/L
glycerol solution, vesicles started to release their contents in
series of bursts (Fig.~\ref{fig:burst}). The intervals between bursts
increased with time. On the other hand, the difference in refractive
indices of the vesicle interior and its exterior decreased. We
attribute these bursts to the the fact that the membrane permeability
for glycerol in the vesicle exterior greatly exceeds the permeability
for the sucrose in the vesicle interior, which results in osmotic
inflating of the vesicle and increasing the membrane tension to the
point where it breaks and releases some of its interior.

Repetitive osmotic swelling followed by a burst-like release has
already been observed in cases where the membrane permeability was
altered by incorporating pore-forming peptides into the membrane; a
theoretical model has also been provided
\cite{Mally:2002,Mally:2007}. Such behaviour has already been
theoretically predicted earlier \cite{Koslov:1984}. In this treatment,
we follow the model proposed in \cite{Mally:2002} with some minor
modifications.

Before going into details, let us summarize the basic ideas of the
``repetitive burst'' model. An impermeable solute is trapped inside a
spherical vesicle, surrounded by a solution of a permeable
solute. Since the concentration of the permeable solute is lower
inside the vesicle, there is a influx of the permeable solute into the
vesicle, accompanied by an influx of water required to maintan the
osmotic balance. During the ``rising'' part of the cycle, the partial
concentration of the impermeable solute is constant, while the partial
concentration of the permeable solute increases, and consequently the
vesicle volume increases as well. During an instantaneous event of a
vesicle burst, some if its content is released. The burst is short
enough that the influx of the permeable solute during the burst is
neglected. During the burst (the ``falling'' phase), the concentration
of either solute inside the vesicle stays constant; however, the total
\emph{quantity} of both the impermeable and the permeable solute
decreases, its decrease being proportional to the decrease of the
vesicle volume.

The osmotic pressure causes an influx of water into an initially
relaxed spherical vesicle with a radius $r_0$. The swelling of a
vesicle strains its membrane and increases its stretching elastic
energy \cite{Evans:MechanicsThermodynamics:1980}:
\begin{equation}
  W = \frac{1}{2} k_a \frac{(A-A_0)^2}{A_0} \; .
  \label{eq:expansion-energy}
\end{equation}
Here, $k_a$ is the area expansivity constant, $A=4\pi r^2$ is the
vesicle membrane area and $A_0=4\pi r_0^2$ the area of an unstrained
vesicle. Equating (\ref{eq:expansion-energy}) with a requirement for a
mechanical equilibrium, $dW = p\,dV$, one obtains an implicit
dependence of the vesicle radius $r$ on the pressure difference $p$
between the inner and outer pressure:
\begin{equation}
  p(r) = \frac{2 k_a (r^2 - r_0^2)}{r_0^2 r} \; .
  \label{eq:pressure-radius}
\end{equation}
As the membrane permeability for water is high enough
\cite{Olbrich:2000} to allow for an almost instantaneous establishment
of osmotic equilibrium, one can equate (\ref{eq:pressure-radius}) with
the osmotic pressure difference $\Pi=kT\Delta c$, thus obtaining
\begin{equation}
  p(r) = kT (c_s + c_g - c_{g0}) \; .
  \label{eq:pressure-concentration}
\end{equation}
Here, $k$ is the Boltzmann constant, $T$ the temperature, and $c_s$,
$c_g$, and $c_{g0}$ the concentrations of sucrose inside the vesicle,
and glycerol inside and outside the vesicle, respectively. As the
experimental chamber can be considered as an infinite reservoir
compared to the volume of a vesicle, $c_{g0}$ can be considered
constant, while $c_g = N_g/V$ and $c_s = N_s/V$ vary with time. Here,
$N_g$ and $N_s$ are the numbers of glycerol and sucrose molecules
inside the vesicle with a volume $V$ at a given time. From the
relation (\ref{eq:pressure-concentration}), one can compute the vesicle
radius $r$ as a function of the number of glycerol molecules inside
the vesicle.

The flux of glycerol molecules into the vesicle is proportional to the
difference of glycerol concentration outside and inside the vesicle,
\begin{equation}
  j = P (c_{g0} - c_g) \; ,
  \label{eq:flux-glycerol}
\end{equation}
where $P$ is the permeability of the membrane for glycerol. The number
of glycerol molecules inside the vesicle increases with time at a
rate:
\begin{equation}
  \frac{dN_g}{dt} = PA \left( c_{g0} - \frac{N_g}{V}\right) \; .
  \label{eq:glycero-rate}
\end{equation}
It is important to stress here that in (\ref{eq:glycero-rate}), both
the vesicle volume $V$ and the membrane area $A$ vary with time, while
the membrane permeability $P$ is constant. At this point the treatment
departs from \cite{Mally:2002}, where the product $PA$ was considered
constant. Using the vesicle radius which can be computed from
(\ref{eq:pressure-concentration}) and inserting it into
(\ref{eq:glycero-rate}), one can solve (\ref{eq:glycero-rate})
numerically.

\begin{figure}
  \centering\includegraphics[scale=0.7]{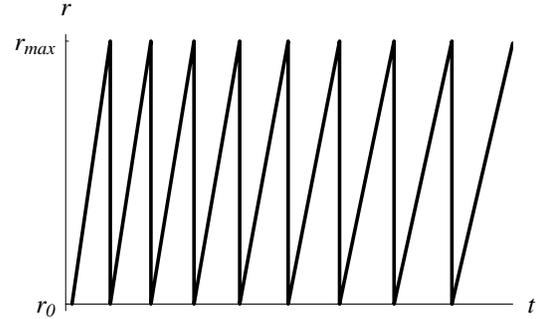}
  \caption{Schematic behaviour of the vesicle radius $r$ as a function
    of time $t$ upon a transfer from a sucrose/glucose solution into
    an iso-osmolar glycerol solution, as assumed by the theoretical
    model.}
  \label{fig:sawtooth}
\end{figure}

For realistic parameter values, however, numerical integration is not
needed. It is known \cite{Bloom:1991} that the lipid membrane can
expand by approximately 4\% before its tensile strength is reached, at
which point the vesicle rapidly releases its excess volume, and its
radius $r$ returns to its equilibrium value $r_0$
(Fig.~\ref{fig:sawtooth}). After the burst, the vesicle begins to
swell again, until the membrane tension reaches its limiting value,
when the vesicle bursts again. In the interval between bursts, the
vesicle volume $V$ and the number of glycerol molecules inside the
vesicle $N_g$ increase. During a burst, part of the sucrose/glycerol
solution inside a vesicle is released in such a way that the
concentrations $c_g$ and $c_s$ do not change.

In the interval between subsequent bursts, the vesicle radius $r$ can be
expanded around $r_0$ and linearized, thus obtaining a simplified
expression relating the change in volume $\Delta V$ with the change of
the number of glycerol molecules which entered the vesicle since the
last burst ($\Delta N_g$):
\begin{equation}
  \Delta V = \frac{\Delta N_g}{\frac{4 k_a}{3 k T r_0} + c_{g0}} 
  \approx \frac{\Delta N_g}{c_{g0}} \; .
  \label{eq:delta-volume}
\end{equation}
The simplification took into account $4k_a/3kTr_0 \ll c_{g0}$. The
number of glycerol molecules that enter the vesicle in a time between
successive bursts ($\Delta N_g^\mathrm{max}$) corresponds to the
maximal increase in the vesicle volume $\Delta V_\mathrm{max}$ and is
thus constant. The number of sucrose molecules ejected in the $i$-th
burst ($\Delta N_s^{(i)}$) is proportional to the number of molecules
present in the vesicle just before the burst ($N_s^{(i)}$):
\begin{equation}
  \Delta N_s^{(i)} = \frac{\Delta V_\mathrm{max}}{V_\mathrm{max}}
  N_s^{(i)} \; .
  \label{eq:delta-sucrose}
\end{equation}

Equation (\ref{eq:glycero-rate}) can be linearized if the vesicle
radius $r$ (and subsequently $A$ and $V$) is expanded around $r_0$,
leading to an expression for the time interval $\Delta t_i$ between
successive bursts:
\begin{equation}
  \Delta t_i = -\frac{\tau}{b} \ln\left(1 - \frac{b\, \Delta
    N_g^\mathrm{max}}{N_s^{(i)}} \right) \; ,
  \label{eq:delta-time-exact}
\end{equation}
where
\begin{eqnarray}
  \tau &=& \frac{V_0}{P A_0} \; , \label{eq:tau} \\
  b &=& \frac{1}{3}\frac{N_s^{(i)}}{c_{g0}V_0} \; . \label{eq:b}
\end{eqnarray}
Linearizing the logarithm in (\ref{eq:delta-time-exact}) yields:
\begin{equation}
  \Delta t_i \approx \frac{r_0}{3P}\frac{c_{g0}\Delta
    V_\mathrm{max}}{N_s^{(i)}} \; .
  \label{eq:delta-time-linearized}
\end{equation}
The expression obtained allows for an estimate of the membrane
permeability $P$ from a videomicroscopy recording of a series of
vesicle bursts:
\begin{equation}
  P = \left\langle \frac{r_0}{3\,\Delta t_i} \frac{c_{g0}\Delta
    V_\mathrm{max}}{N_s^{(i)}} \right\rangle_i \; .
  \label{eq:permeability}
\end{equation}
Here, $r_0$ and $\Delta V_\mathrm{max}$ can be determined directly
from a microscope image, $c_{g0}$ is known, $\Delta t_i$ can be
measured, and $N_s^{(i)}$ can be determined from the rate at which the
vesicle halo decreases in intensity \cite{Mally:2002,Mally:2007},
given that its initial value is known, $N_s^{(0)}=c_s^{(0)} V_0$. The
average runs over a series of $i$ recorded bursts. Calculated values
for membrane permeability for five recorded series of bursts are
collected in Table~\ref{tab:permeability}, yielding an average value
$P = (2.09\pm0.82)\cdot 10^{-8}\,\textrm{m/s}$ at room temperature
($26\pm2^\circ$C). Within the experimental error, this value fits the
published value for DOPC, $P = 2.75\cdot 10^{-8}\,\textrm{m/s}$ at
$30^\circ$C \cite{Paula:1996}. Earlier estimates \cite{Orbach:1980},
however, were approximately twice as high ($5.4\cdot
10^{-8}\,\textrm{m/s}$ at $25^\circ$C).

\begin{table}
  \caption{Permeability of the POPC membrane for glycerol, calculated from
    (\protect\ref{eq:permeability}) for five recorded transfers of
    the POPC vesicle into an iso-osmolar solution of glycerol.}
  \label{tab:permeability}
  \begin{tabular}{l@{\hspace{2em}}l@{\hspace{2em}}r}
  \hline
  experiment & \# of bursts & $P$ $[10^{-8}\;\textrm{m/s}]$ \\
  \hline
  1 & 24 & $1.81 \pm 0.84$ \\
  2 & 30 & $2.09 \pm 1.07$ \\
  3 & 31 & $2.19 \pm 0.66$ \\
  4 & 22 & $2.03 \pm 0.73$ \\
  5 & 21 & $2.33 \pm 0.62$ \\
  \hline
  \end{tabular}
\end{table}

No bursts were observed when the vesicles were still attached to the
electrodes and the external solution was exchanged.  On one hand, this
is in agreement with the solution exchange mechanism suggested by
Estes and Mayer, yet on the other hand this does not disprove the
model of solute exchange proposed in the paper.  As long as the
vesicle is connected to the substrate with a tether, the area of the
vesicle membrane is not a well-defined quantity, since new lipid
material can be incorporated into the membrane in order to accomodate
an increase of the vesicle volume.

No vesicle bursts were observed upon a micropipette transfer of a
vesicle from a sucrose solution to an iso-osmolar glucose
solution. This leads us to believe that the key factor for repetitive
bursts is the mismatch in the permeabilities of phospholipid membrane
for glycerol on the one hand and for either sucrose or glucose on the
other.

\subsection{Preparing flaccid vesicles}

As opposed to spontaneous swelling methods
\cite{Reeves:1969,Needham:1988}, which usually yield a variety of
vesicles shapes, both spherical and flaccid, electroformation is
generally considered as a method for producing spherical vesicles
\cite{Dimova:2006}.  If flaccid vesicle shapes are required, one can
leave the experimental chamber open for several hours (J.~Majhenc,
personal communication): as the concentration of the external solution
increases due to evaporation, osmotic pressure deflates the vesicle.
The procedure is very time-consuming and gives the experimentalist
poor control over the intended relative volume of the vesicle.

Here, we propose a method for flaccid vesicle preparation using
electroformation which relies on relatively high permeability of the
phospholipid membrane for glycerol.  The vesicle is initially filled
with a mixture of aqueous solutions of a permeable and an impermeable
solute, in our case glycerol and sucrose.  As long as the composition
of the external solution stays the same, this induces no changes of
the vesicle shape, as the permeable solute traverses the membrane in
equal amount inwards and outwards.  Once the composition of the
external iso-osmolar solution changes to a lower concentration of the
permeable solute, the permeable solute starts to diffuse outwards
through the membrane until the concentrations of the permeable solute
equilibrates on both sides.  In order to maintain the osmotic balance,
the water leaves the vesicle as well, thus deflating the vesicle.  To
illustrate, a vesicle filled with 70\% of an impermeable solute
(sucrose) and 30\% of an permeable solute (glycerol) is expected to
deflate to 70\% of its initial volume once the external solution
changes to an iso-osmolar solution of the impermeable solute
(glucose).  A sensible parameter which characterizes the shape of the
vesicle regardless of its size is the relative volume $v$, which is
defined for a vesicle with a volume $V$ and a membrane area $A$ as the
volume $V$, divided by the volume of a sphere with the area $A$; $v$
can attain values from 0 to 1.  In the above example, the relative
volume of the vesicle is expected to change from 1.0 to 0.7.

Two kinds of experiments were conducted.  In the first experiment, the
electroformation chamber was initially filled with a mixture of
glycerol and sucrose solutions.  The ratios from 70:30 v/v to 90:10
v/v of 0.2~mol/L sucrose and 0.2~mol/L glycerol solutions were tried.
After the vesicle were formed, the solution in the electroformation
chamber was exchanged with 0.2~mol/L glucose.  Although the relative
volume of the drained vesicles has not been precisely measured,
\emph{e.g.}, with sucking the vesicle into a micropipette or by its
deformation in an applied electric field, it was estimated that it was
significantly higher than the value which follows from the model.
Another problem we encountered was that a majority (estimated at over
90\%) of the vesicle obtained by this method was spherical with
spherical invaginations rather than flaccid, indicating that a change
in the vesicle volume has not been accompanied by a corresponding
change in the difference between the outer and the inner membrane
leaflet.  We believe that the systematically higher relative volume
can be related to the detachment stress when the vesicles were teared
from the substrate during the draining.

\begin{figure}
  \centering\includegraphics[scale=0.55]{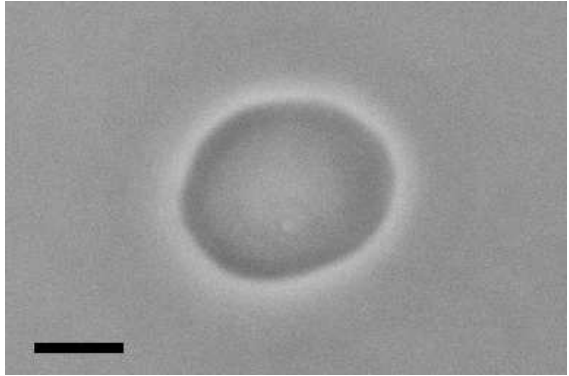}
  \caption{A flaccid POPC vesicle prepared using electroformation in a
    90:10 v/v 0.2~mol/L sucrose, 0.2~mol/L glycerol solution, drained
    and flushed with an equal volume of a 0.2~mol/L glucose solution.
    Expected relative volume $v=0.95$.  The bar represents 20~$\mu$m.}
  \label{fig:flaccid}
\end{figure}

In another experiment, a procedure similar to the preparation of
vesicles in a sucrose/glucose solution (described in
section~\ref{sec:procedure}) was used.  Vesicles were prepared using
the electroformation in a mixture of 0.2~mol/L sucrose and 0.2~mol/L
glycerol, ranging from 70:30 v/v to 90:10 v/v.  After the
electroformation procedure was completed, the chamber was drained and
immediately flushed with an iso-osmolar solution of glucose, equal in
volume to the volume of the electroformation chamber and the tubing.
After dilution, the concentration of glycerol in the external solution
dropped to a half of its initial value.  Therefore, we expected a
change in the relative volume which corresponds to the difference in
the glycerol concentration, \emph{i.e.}, a 70:30 mixture is expected
to yield vesicles with $v=0.85$.  As in the above experiment, flaccid
vesicles were outnumbered by vesicles with either in- or evaginations.
The observed relative volumes were systematically higher in the 70:30
mixture, while more consistent results were obtained in the 90:10
mixture (Fig.~\ref{fig:flaccid}).  With varying the initial and the
final concentration of glycerol in the mixture and/or the volume of
the solution with which the electroformation chamber is flushed, a
more precise control of vesicle volume in the range $0.9 < v < 1.0$
ought to be possible.

\section{Discussion}
\label{sec:discussion}

\subsection{Estimating the rate of vesicle interior exchange}

Using a quick estimate we can try to verify whether the mechanism
proposed by Estes and Mayer (\cite{Estes:2005b}, Fig. 6) can fully
explain the rapid exchange of glycerol in the vesicle interior with
the NaCl from the exterior solution. As an example, we consider a
vesicle with a 100~$\mu$m diameter ($R=50\;\mu\textrm{m}$), attached
to a tether with the length $L$. For simplicity we limit our treatment
to the diffusion of glycerol, since its mobility in an aqueous
solution is smaller than the mobilities of sodium or chlorine ions,
and is thus the rate-limiting factor. If we approximate the glycerol
concentration change in the vesicle exterior with a step-wise drop,
the concentration of glycerol inside the vesicle would follow
exponentially:
\[ c(t) = c_0 e^{-t/\tau} \; .
\]
Here, $\tau$ is the characteristic time, which can be expressed as a
product of the vesicle volume $V$ and the ``diffusion resistance'' of
the tether $R_d$.  Here, we treated the vesicle exterior as an
infinite reservoir compared to its interior.  The latter can in turn
be expressed as $R_d = L/(SD)$, $S$ being the cross-section of the
tether, and $D$ the diffusion coefficient for glycerol in water.  The
diffusion coefficient depends strongly on the temperature and the
glycerol concentration, as a rough estimate, the value $D=1.0\cdot
10^{-5}$~cm${}^2$/s will be taken \cite{Ternstrom:1996,DErrico:2004}.
Taking into account the above mentioned value for $R$ and $L =
30\;\mu\textrm{m}$ (the longest tether length reported by Estes and
Mayer), and making an estimate $S=1$~$\mu\mathrm{m}^2$ (which is
likely an overestimate), one obtains $\tau\sim 1.5\cdot 10^4$~s, or
approximately 4.5 hours, which exceeds the observed changes on the
timescale of minutes by two orders of magnitude.  Only the exchange
through very short tethers ($L = 1\;\mu\textrm{m}$) can be explained
by this mechanism, yielding $\tau\sim 500$~s, or approximately
8~minutes.

Instead, we believe it is necessary to consider the permeability of
the phospholipid membrane for glycerol.  Carrying out a similar
estimate for the membrane permeation, we obtain $\tau = R_d V =
V/(AP)$, with $A$ being the vesicle membrane area and $P = 2.09\cdot
10^{-6}\;\textrm{cm/s}$ the permeability of the phospholipid membrane
for glycerol, yielding $\tau \approx 800$~s, which is close to the
experimentally observed values and comparable with the exchange
through very short tethers.  While the exchange of glycerol was
considered rate-limiting during the solute transport through the
tether tube, the situation is reversed here.  The membrane is much
more permeable to glycerol than to anions (permeability of DOPC
bilayer for $\textrm{Cl}^-$ is $1.2\cdot 10^{-8}\;\textrm{cm/s}$,
\cite{Paula:1998}), and the membrane permeability to cations is even
lower \cite{Deamer:1986}.  Even lower is the membrane permeability for
sugars (permeability of DMPC bilayer for glucose is $1.4\cdot
10^{-10}\;\textrm{cm/s}$, \cite{Bresseleers:1984}).  The latter fact
explains why the compositional gradient of vesicles filled with
sucrose and immersed in a iso-osmolar sucrose/glucose medium can be
preserved for days, yet it also implies that the transport of sugars
or ions through the membrane is negligible.  This, as well as the fact
that Estes and Mayer report that the exchange occured more rapidly in
the vesicles closest to the substrate, led us to believe that both
pathways contribute to the solute exchange in the vesicle interior.

\subsection{Free vs. attached vesicles}

While it was customary in earlier works (\emph{e.g.},
\cite{Angelova:1992}) to perform experiments in the same chamber where
the electroformation was done, researchers have lately become aware of
the fact that the vesicles in the electroformation chamber are
attached to the substrate as a rule \cite{Dimova:2006}, and thus a
comparision with theoretical models valid for free vesicles is
doubtful. Instead, they adopted a technique where the vesicles are
drained from the electroformation chamber and stored separately. In
addition, this offers a possibility to dilute the vesicle suspension
from the electroformation chamber with another iso-osmolar solution; a
popular combination is electroformation in a sucrose solution and
dilution with a glucose solution \cite{Mally:2002,Riske:2005}, which
allows easier experimentation due to the greater specific weight of
vesicles and their easier discrimination from the background in a
phase-contrast setup.

\begin{figure}
  \centering\includegraphics[scale=0.65]{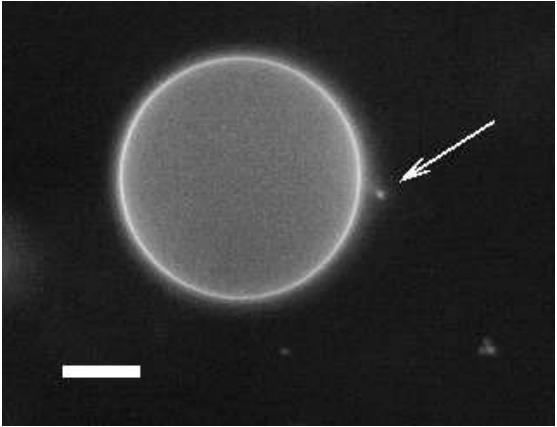}
  \caption{Tether residues (arrow) are still visible on vesicles
    minutes after draining from the electroformation chamber, as
    evident on this POPC vesicle labelled with 1 wt\%~NBD. The bar
    represents 10~$\mu$m.}
  \label{fig:tether-fluoro}
\end{figure}

Tethers with which the vesicles are attached to the substrate are torn
in the process.  Tether residues can be directly visualized. Minutes
after draining, small tubular protrusions can still be seen
(Fig.~\ref{fig:tether-fluoro}) on vesicles, which we interpret as
remains of tethers with which the vesicles were attached to the
substrate.  While the re-sealing of the tether on the spot where it
was torn away from the substrate probably occurs promptly, the
integration of the residual tubular structure into the spherical
mother vesicle using relaxation processes usually takes hours
\cite{KraljIglic:2001a}.  A theoretical model has been proposed to
explain this phenomenon \cite{Bozic:2002}.

\subsection{Metal electrode vs. ITO glass as a substrate}

There are two variants of the electroformation in use, both pioneered
by Angelova and coworkers.  The first one uses a pair of parallel
platinum wires as electrodes \cite{Angelova:1986}, while the second
one utilizes a pair of plan-parallel ITO-coated glasses
\cite{Angelova:1992}.  The latter method has one obvious advantage:
since it exposes a greater surface on which the vesicles can grow, it
provides a higher yield, especially if techniques for the uniform
spreading of the lipid film are used \cite{Estes:2005a}.  In cases
where the overall yield is low, \emph{e.g.}, when preparing GUVs in a
conductive medium, this advantage can be crucial.  Among its
disadvantages, one has to mention that ITO-coated glasses can be
damaged easily by the Joule heat of the applied electric current,
which limits the solution conductivities below those achievable by the
setup with platinum electrodes.  Another disadvantage is that the
optical axis is aligned with the direction of the electric field,
which sometimes makes effects like aggregation on the electrodes
harder to observe.

\section{Conclusions}

In this paper, we examined a recently reported method
\cite{Estes:2005b} for producing phospholipid GUVs using
electroformation, where the solution in the electroformation chamber
vesicles is replaced by another solution after or during the
electroformation process. By testing the system with different initial
and final solutions, we noticed that the exchange of the vesicle
interior, as reported by Estes and Mayer, has only been observed in
the case where either the initial or the final external solution was a
solution of glycerol. That has led us to think that the exchange of
vesicle interior can be at least in part attributed to a relatively
high permeability of the phospholipid membrane for glycerol rather
than being explained solely by the exchange through the tethers with
which the vesicles are connected to the substrate.

To test this hypothesis, the permeability of the POPC membrane for
glycerol was measured in an experiment where a single GUV was
transferred using a micropipette from a 0.2~mol/L sucrose/glucose
solution into an iso-osmolar glycerol solution. Videomicroscopy
monitoring showed that upon transfer, vesicle radius started to
increase until the point where the membrane tensile strength was
reached, whereupon the vesicle ejected part of its volume in a burst
and its radius returned to its initial value, and the process was
repeated over again. We attributed this behaviour to high membrane
permeability for glycerol and calculated the permeability of the POPC
membrane for glycerol from the observed parameters. The calculated
value for permeability lies within the experimental error of an
already published value obtained by a different method
\cite{Paula:1996}, and is approximately $10^4\times$ higher than the
published value for the membrane permeability for sugars
\cite{Bresseleers:1984}, which confirms our initial hypothesis.

In addition, we have demonstrated that a high membrane permeability
for glycerol can be utilized for preparing flaccid vesicles, where
electroformation was conducted in a mixed sucrose/glycerol solution
and later exchanged with an iso-osmolar glucose solution.  This shows
that the electroformation method in a flow chamber introduced by Estes
and Mayer \cite{Estes:2005b} is a powerful method which can be
deployed to many different situation; among those, the preparation of
GUVs in a pH gradient \cite{Hope:1989,Redelmeier:1990,Mathivet:1996},
which opens way to new experiments.

\appendix

\section{Estimating the field in the electroformation chamber}
\label{sec:bipolar}

\begin{figure}
  \centering\includegraphics[scale=0.8]{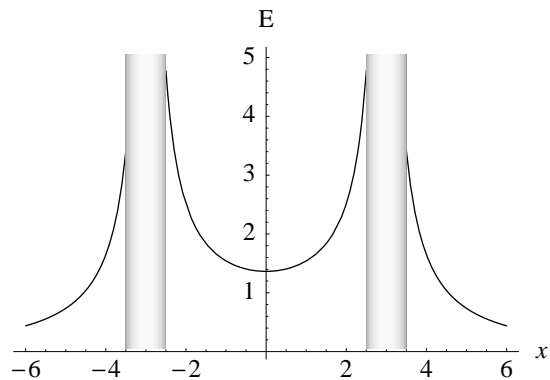}
  \caption{Electric field along the axis connecting the centres of
    both cylindrical electrodes inside the electroformation
    chamber. Electric field $E$ is expressed in term of the field in
    the plate capacitor, $V_0/d$, and the spatial coordinate $x$ is in
    milimetres. The parameters $R=0.5$, $d=5$ were used in the
    calculation.}
  \label{fig:bipolar-el-field}
\end{figure}

An estimate of the electric field in the electroformation chamber with
the geometry shown in Fig.~\ref{fig:electroform-chamber} can be most
easily obtained in the bipolar coordinate syetem
\cite{MorseFeshbach:MethTheorPhys:p1210}. The relation between the
bipolar and the Cartesian coordinates:
\begin{eqnarray}
  x &=& \frac{a \sinh\xi}{\cosh\xi + \cos\theta} \; , \\
  y &=& \frac{a \sin\theta}{\cosh\xi + \cos\theta} \; .
\end{eqnarray}
Equations $\xi =\textrm{const.}$ represent circles with the foci at
$(0,-a)$ or $(0,a)$. The coordinate $\xi$ is ``radial'', with
$\xi=-\infty$ in $(0,-a)$ and $\xi=+\infty$ in $(0,a)$, while the
coordinate $\theta$ is an angular one, $\theta \in [0,2\pi]$.  The
surface of the cylindrical electrode corresponds to a certain value of
$\xi$, here denoted by $\xi_0$ for positive values of $\xi$ ($-\xi_0$
corresponds to the electrode described by negative values of $\xi$).
The parameters $a$ and $\xi_0$ can be related to the electrode radius
$R$ and the distance $d$ between the electrodes:
\begin{eqnarray}
  R &=& \frac{a}{\sinh\xi_0} \; , \\
  d &=& 2R(\cosh\xi_0 -1) \; .
\end{eqnarray}
The advantage of the chosen coordinate system is that the Laplace
equation for the electric potential $V$ separates in the given
geometry, when expressed in terms of $(\xi,\theta)$. Requiring
$V(\xi_0)=V_0$, $V(-\xi_0)=-V_0$, one obtains:
\begin{equation}
  V(\xi) = V_0 \frac{\xi}{\xi_0} \; .
\end{equation}
Hence the field
\begin{eqnarray}
  E_\xi &=& -\frac{1}{h_\xi} \frac{\partial V}{\partial\xi} = \nonumber \\
  &=& - \frac{2(\cosh\xi_0-1)}{\xi_0\sinh\xi_0} \frac{V_0}{d} 
  (\cosh\xi+\cos\theta)\; ,
\end{eqnarray}
with $h_\xi=a/(\cosh\xi+\cos\theta)$ being the corresponding metrical
coefficient. Fig.~\ref{fig:bipolar-el-field} depicts the calculated
electric field between two parallel cylinders in an unconfined space
filled with a homogeneous medium.  On the axis connecting the centres
of the two electrodes, the electric field on the outer side is
approximately 71\% of its value on the inner side.

For a more realistic model of the field inside the electroformation
chamber, one would need to take into account the proximity of chamber
walls, which contribute to an increase of the field between the
electrodes, since glass is less conductive than the aqueous medium.
Another factor that affects the field in the electroformation chamber
is the layer of lipid adsorbed to the electrodes, which decreases the
field in the aqueous medium, as most of the voltage drop occurs in the
lipid layer.  In particular for the assesment of the effect of the
geometrical confinement, numerical modelling is required.

\subsection*{Acknowledgement}

The authors thank Dr. Janja Majhenc and Dr. Mojca Mally for helpful
discussions.  This work has been supported by the Slovenian Research
Agency grant P1-0055.








\end{document}